\documentstyle[12pt,axodraw]{article}
\newcommand{\calA}{{\cal A}}
\newcommand{\calL}{{\cal L}}
\newcommand{\calU}{{\cal U}}

\newcommand{\oP}{\textrm{o-Ps}}
\newcommand{\pP}{\textrm{p-Ps}}
\newcommand{\Br}{\textrm{Br}}

\begin{document}
\baselineskip=16pt

\pagenumbering{arabic}

\vspace{1.0cm}

\begin{center}
{\Large\sf Bounds on unparticles couplings to electrons: from
electron $g-2$ to positronium decays}
\\[10pt]
\vspace{.5 cm}

{Yi Liao\footnote{liaoy@nankai.edu.cn}}

{\small Department of Physics, Nankai University, Tianjin 300071,
China} %

\vspace{2.0ex}

{\bf Abstract}

\end{center}

Unparticles as suggested recently by Georgi have surprising
phenomenological implications, distinctive from any other new
physics that we know of. But they must interact very feebly with
ordinary matter to have avoided detection thus far. We determine
how feebly they can interact with the electron, using the
precisely measured quantities in QED: the electron $g-2$ and the
bounds on invisible and exotic positronium decays. The most
stringent bound comes from invisible orthopositronium decays: the
effective energy scale entering the vector unparticle-electron
interaction must exceed $4\times 10^5$ TeV for a scaling dimension
$\frac{3}{2}$ of the vector unparticle. The lower bounds on scales
for other unparticles range from a few tens to a few hundreds TeV.
This makes the detection of unparticles challenging in low energy
electron systems.

\begin{flushleft}

\end{flushleft}

\newpage
We are so accustomed to describe physical processes in terms of
particles that it is even hard to imagine what other conception we
can perceive beyond that of particles. By particles we mean
identities that have a definite energy-momentum relation, i.e., a
mass, among other intrinsic properties. Recently, Georgi has
suggested a fascinating idea of what this beyond-particle
identity, dubbed unparticle, might look like
\cite{georgia,georgib}. He has also provided a scenario in which
the unparticle could appear and couple to ordinary matter from
certain high energy theory with a nontrivial scale invariant
infrared fixed point, for instance, theories studied in
Ref.\cite{banks}. Although not much is known about the details of
such a high energy theory that might be relevant to the real
world, its remnants at low energies, unparticles, can be well
described in effective field theories and experimentally explored
through their couplings to ordinary matter. As Georgi argued and
demonstrated \cite{georgia,georgib}, these unparticles enjoy very
funny kinematic behavior, far removed from any new physics that we
know of so far. This makes the idea phenomenologically attractive.

However peculiar these unparticles might be, they must interact
very feebly with ordinary matter to evade detection so far. It is
the aim of the current work to determine how feeble those
interactions might be from two of the most precisely measured
quantities in QED: the electron $g-2$ and the invisible and exotic
decays of positronium. For an unparticle of scaling dimension
$\frac{3}{2}$, we find that the former restricts the effective
energy scale responsible for the unparticle-electron interactions
to be higher than tens to $150$ TeV, depending on the Lorentz
properties of the unparticle. The constraint from positronium
decays is more stringent: the lower bound ranges from $500$ to
$4\times 10^5$ TeV.

Prior to this work, three papers on unparticles phenomenology
appeared, but no systematic analysis has been attempted so far on
experimental constraints on unparticle-electron interactions. In
Ref.\cite{georgib}, unparticle effects at the $Z$ resonance were
elucidated where unusual patterns of interference occur due to the
phases in the unparticle propagator in the time-like region. In
Ref.\cite{cheung}, based on effective operators suggested in
Ref.\cite{georgia}, collider signals of unparticles are studied
together with effects of a vector unparticle on the muon $g-2$.
The idea of bosonic unparticles was generalized to the fermionic
case in Ref.\cite{luo}, where corrections of fermionic and scalar
unparticles to the muon $g-2$ are computed as well as potential
flavor-changing neutral current effects mediated by a vector
unparticle.

Our working Lagrangian for effective unparticle-electron
interactions is
\begin{eqnarray}
\calL_{\textrm{int}}=
C_S\overline{\psi}\psi\calU_S+C_P\overline{\psi}i\gamma_5\psi\calU_P
+C_V\overline{\psi}\gamma_{\mu}\psi\calU^{\mu}_V
+C_A\overline{\psi}\gamma_{\mu}\gamma_5\psi\calU^{\mu}_A,
\end{eqnarray}
where $\calU_{S,P,V,A}$ are fields for scalar, pseudoscalar,
vector and axial vector unparticles. They have standard $C$ and
$P$ parities as their particle counterparts to preserve $C$ and
$P$ symmetries. Although these fields may have different scaling
dimensions, we assign a common one to them for simplicity, $d$.
The real couplings $C_{S,P,V,A}$ then have the dimension $1-d$ and
can be parametrized by $C_{S,P,V,A}=\pm\Lambda_{S,P,V,A}^{1-d}$,
where $\Lambda_i$ are effective energy scales determined by some
underlying high energy theory. Our goal is to constrain these
energy scales using the precisely measured electron $g-2$ and the
upper limits on invisible and exotic decays of the positronium.

By exploiting scale invariance of the unparticle field, Georgi
found that the state density in phase space of an unparticle of
momentum $p$ is proportional to
$\theta(p^0)\theta(p^2)(p^2)^{d-2}$. Since these are the same
factors for the density of a system of $d$ massless particles, he
suggested that the state density of an unparticle is similarly
normalized,
\begin{eqnarray}
d\Phi_{\calU}(p)=A_d\theta(p^0)\theta(p^2)(p^2)^{d-2}
\frac{d^4p}{(2\pi)^4},
\end{eqnarray}
where
\begin{eqnarray}
A_d=\frac{16\pi^{\frac{5}{2}}}{(2\pi)^{2d}}
\frac{\Gamma(d+\frac{1}{2})}{\Gamma(d-1)\Gamma(2d)},
\end{eqnarray}
though $d$ is now generally nonintegral. This should be contrasted
to that of a particle of mass $m$:
\begin{eqnarray}
d\Phi(p)=2\pi\theta(p^0)\delta(p^2-m^2)\frac{d^4p}{(2\pi)^4}.
\end{eqnarray}
Note that there is no mass-shell constraint to an unparticle, as
is the case for a particle. This will have interesting
phenomenological implications. From unitarity considerations, the
above state density implies the following propagator for a
spin-zero unparticle \cite{georgib} (see also \cite{cheung}):
\begin{eqnarray}
\frac{A_d}{2\sin(\pi d)}\frac{i}{(-p^2-i\epsilon)^{2-d}}.
\end{eqnarray}
For a vector or axial vector unparticle that has only transverse
polarizations, a standard projector should be attached,
$P^T_{\mu\nu}=-g_{\mu\nu}+p_{\mu}p_{\nu}/p^2$.

\begin{figure}
\begin{picture}(400,80)(0,0)
\SetOffset(20,10)
\ArrowLine(10,10)(60,40)\ArrowLine(60,40)(10,70)%
\Photon(60,40)(80,40){1.5}{3}%
\DashLine(20,18)(20,62){3}\DashLine(21,18)(21,62){3}
\Text(65,30)[l]{{\small$q,\mu$}}\Text(8,10)[r]{{\small$p$}}
\Text(45,0)[]{{\small(a)}}

\SetOffset(120,10)
\ArrowLine(0,10)(40,40)\ArrowLine(40,40)(0,70)%
\DashLine(40,39.5)(70,39.5){3}\DashLine(40,40.5)(70,40.5){3}
\Text(60,32)[l]{{\small$p$}}%
\Text(20,18)[]{{\small$p_1$}}\Text(20,62)[]{{\small$-p_2$}}
\Text(35,0)[]{{\small(b)}}

\SetOffset(220,10)
\ArrowLine(0,10)(30,10)\ArrowLine(30,10)(30,70)\ArrowLine(30,70)(0,70)%
\DashLine(30,9.5)(60,9.5){3}\DashLine(30,10.5)(60,10.5){3}
\Photon(30,70)(60,70){1.5}{4}
\Text(45,15)[l]{{\small$p$}}\Text(45,62)[]{{\small$k,\mu$}}
\Text(15,15)[]{{\small$p_1$}}\Text(15,62)[]{{\small$-p_2$}}
\Text(80,0)[]{{\small(c)}}

\SetOffset(320,10)
\ArrowLine(0,10)(30,10)\ArrowLine(30,10)(30,70)\ArrowLine(30,70)(0,70)%
\DashLine(30,69.5)(60,69.5){3}\DashLine(30,70.5)(60,70.5){3}
\Photon(30,10)(60,10){1.5}{4}
\Text(45,62)[l]{{\small$p$}}\Text(45,15)[]{{\small$k,\mu$}}
\Text(15,15)[]{{\small$p_1$}}\Text(15,62)[]{{\small$-p_2$}}
\end{picture}
\caption{Diagrams that contribute to electron $g-2$ (a), invisible
(b) and exotic (c) positronium decays. Double-dashed, solid and
wavy lines stand for unparticle, electron and photon fields
respectively.}
\end{figure}
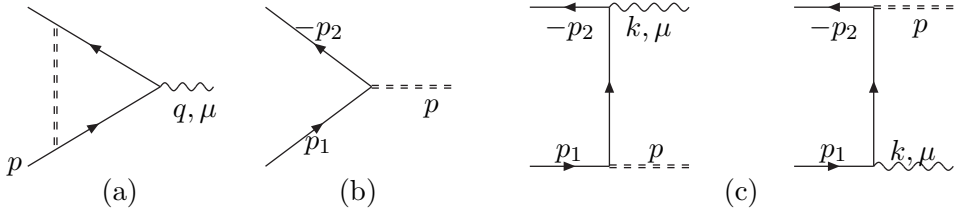

It is straightforward to work out corrections to the anomalous
magnetic moment of the electron, $a=\frac{1}{2}(g-2)$, from Fig.
1(a):
\begin{eqnarray}
a_S&=&-\frac{A_d}{2\sin(\pi d)}\frac{(C_Sm^{d-1})^2}{8\pi^2}
\frac{3\Gamma(2d-1)\Gamma(2-d)}{\Gamma(2+d)},\\
a_P&=&+\frac{A_d}{2\sin(\pi d)}\frac{(C_Pm^{d-1})^2}{8\pi^2}
\frac{\Gamma(2-d)\Gamma(2d)}{\Gamma(2+d)},\\
a_V&=&-\frac{A_d}{2\sin(\pi d)}\frac{(C_Vm^{d-1})^2}{4\pi^2}
\frac{\Gamma(3-d)\Gamma(2d-1)}{\Gamma(d+2)},\\
a_A&=&+\frac{A_d}{2\sin(\pi d)}\frac{(C_Am^{d-1})^2}{\pi^2}
\frac{\Gamma(2d-2)\Gamma(3-d)}{\Gamma(2+d)},
\end{eqnarray}
where the subscripts denote the contributions from corresponding
unparticles. Note that $a_A$ is computed for a transverse
$\calU_A$ while $a_V$ does not rely on the transversality
assumption for $\calU_V$. For the relevant loop integrals to
converge it is necessary that $d<2$. As argued in
Ref.\cite{georgia}, theoretical consistency may demand that $d>1$.
Our later numerical analysis will thus focus on the narrow range
of the scaling dimension, $1<d<2$. It is then clear that
$a_{S,V}>0$ while $a_{P,A}<0$. Our result on $a_V$ coincides with
that in Ref.\cite{cheung}, while $a_S$ differs in sign from
Ref.\cite{luo}. In the limit $d\to 1$, we have
$a_S\to\frac{3C_S^2}{16\pi^2},~a_P\to-\frac{C_S^2}{16\pi^2},~
a_V\to\frac{C_V^2}{8\pi^2}$ while $a_A$ has no appropriate limit
due to infrared divergence. The conventional one-loop QED result
is recovered from $a_V$ by setting further $C_V\to -e$.

The electron $g-2$ has been recently measured in Ref.\cite{odom}
(denoted as H06) with an uncertainty about 6 times smaller than in
the past. Using as input the fine structure constant measured in
independent experiments with Cs \cite{gerginov} (Cs06) and Rb
\cite{clade} (Rb06) atoms, in the new theoretical evaluation of
$g-2$ \cite{kinoshita}, yields the following deviations
\cite{gabrielse} between the theoretical and measured numbers:
\begin{eqnarray}
a(\textrm{Cs06})-a(\textrm{H06})&=&-2.5(9.3)\times 10^{-12},\\
a(\textrm{Rb06})-a(\textrm{H06})&=&+7.9(7.7)\times 10^{-12},
\end{eqnarray}
which are summarized in Ref.\cite{odom} as
\begin{eqnarray}
|\delta a|<15\times 10^{-12}.
\end{eqnarray}
This last bound will be used below to constrain the
unparticle-electron couplings.

For numerical illustration, we assume $d=\frac{3}{2}$, then
\begin{eqnarray}
a_S=\frac{1}{10\pi^3}\frac{m}{\Lambda_S},~
a_P=-\frac{1}{15\pi^3}\frac{m}{\Lambda_P},~
a_V=\frac{1}{30\pi^3}\frac{m}{\Lambda_V},~
a_A=-\frac{2}{15\pi^3}\frac{m}{\Lambda_A}.
\end{eqnarray}
We will not attempt here a sophisticated data fitting; instead, we
assume that only one of the four unparticles exists at a time. The
separate bounds are found to be
\begin{eqnarray}
\Lambda_S>110\textrm{ TeV},~~~\Lambda_P>73\textrm{
TeV},~~~\Lambda_V>37\textrm{ TeV},~~~\Lambda_A>146\textrm{ TeV}.
\end{eqnarray}
If all unparticles are accommodated simultaneously, only a bound
on the combination of $\Lambda$'s can be set which is generally
weaker due to cancellations. We mention in passing that the bounds
become weakened when $d$ increases.

Now we move to the positronium decays. A positronium of orbital
and spin angular momenta $\ell,~s$ has parities
$C=(-1)^{\ell+s},~P=(-1)^{\ell+1}$. Thus, the ground-states have
respectively, $C=P=-1$ for an orthopositronium ($\oP$ with $s=1$)
and $-C=P=-1$ for a parapositronium ($\pP$ with $s=0$). While a
$\pP$ must decay into an even number of photons, an $\oP$ has to
decay into an odd number of photons, at least three. This makes
the latter a particularly sensitive probe for new physics effects.
For obvious reasons, we restrict ourselves to decays involving a
single unparticle in the final state. Then, only the following
invisible one-body transitions are allowed:
\begin{eqnarray}
\oP\to\calU_V;~~~\pP\to\calU_P;
\end{eqnarray}
while for exotic two-body decays, the following ones are possible:
\begin{eqnarray}
\oP\to\calU_S\gamma,~\calU_P\gamma,~\calU_A\gamma;~~~
\pP\to\calU_V\gamma;
\end{eqnarray}
where the last one cannot compete with the dominant two-photon
decay and thus will not be considered below. These symmetry
arguments have been checked against explicit calculations.

The amplitudes for the constituent processes shown in Fig. 1(b)
and Fig. 1(c) are found in the nonrelativistic limit:
\begin{equation}
\begin{array}{rcl}
i\calA(\calU_P)&=&-2mC_P\zeta^{\dagger}\xi,\\
i\calA(\calU_V)&=&+i2mC_V\zeta^{\dagger}\sigma^i\xi\epsilon^{i*}(p),\\
i\calA(\gamma\calU_S)&=&-i2eC_S\zeta^{\dagger}\sigma^i\xi
\epsilon^{i*}(k),\\
i\calA(\gamma\calU_P)&=&-i2eC_P\zeta^{\dagger}\sigma^k\xi
\epsilon^{ijk}\hat{k}^j\epsilon^{i*}(k),\\
i\calA(\gamma\calU_A)&=&-2eC_A\zeta^{\dagger}\sigma^k\xi
\epsilon^{ijk}\epsilon^{i*}(k)\epsilon^{j*}(p),
\end{array}
\end{equation}
where $p,~\epsilon(k)$ ($k,~\epsilon(k)$) are the momentum and
polarization of the photon (unparticle), and $\xi,~\zeta$ are the
spin wave-functions for the electron and positron of mass $m$. The
decay amplitudes for the positronium are
\begin{equation}
\begin{array}{rcl}
i\calA(\pP\to\calU_P)&=&-2\sqrt{2m}C_P\psi(0),\\
i\calA(\oP\to\calU_V)&=&+i2\sqrt{2m}C_V\psi(0){\bf
n}\cdot\epsilon^*(p),\\
i\calA(\oP\to\gamma\calU_S)&=&\displaystyle
-i2\sqrt{\frac{2}{m}}eC_S\psi(0){\bf n}\cdot{\bf\epsilon}^*(k),\\
i\calA(\oP\to\gamma\calU_P)&=&\displaystyle
-i2\sqrt{\frac{2}{m}}eC_P\psi(0)
\left({\bf\epsilon}^*(k)\times\hat{{\bf k}}\right)\cdot{\bf n},\\
i\calA(\oP\to\gamma\calU_A)&=&\displaystyle
-2\sqrt{\frac{2}{m}}eC_A\psi(0)\left(
{\bf\epsilon}^*(k)\times{\bf\epsilon}^*(p)\right)\cdot{\bf n},
\end{array}
\end{equation}
where ${\bf n}$ is the $\oP$ polarization vector and $\psi(0)$ is
the wave-function of the positronium bound state evaluated at the
origin.

Since the polarization dependence is standard, we will study
directly the unpolarized decay rates. Again for simplicity, we
will assume that the vector and axial vector unparticles have only
transverse polarizations. The decay rate to a single unparticle is
\begin{eqnarray}
d\Gamma=\frac{1}{4m}A_d\theta(p^0)\theta(p^2)(p^2)^{d-2}
\frac{d^4p}{(2\pi)^4}(2\pi)^4\delta^4(p-p_1-p_2)|\calA|^2,
\end{eqnarray}
which can be integrated to
\begin{eqnarray}
\Gamma=4^{d-3}m^{2d-5}A_d|\calA|^2.
\end{eqnarray}
Note that in contrast to the particle case there is no delta
function remaining because unparticles have no mass-shell
constraints. The invisible decay rates are
\begin{eqnarray}
\Gamma(\pP\to\calU_P)&=&
2^{2d-3}A_d~m (m^{d-1}C_P)^2|m^{-3/2}\psi(0)|^2,\\
\overline{\Gamma}(\oP\to\calU_V)&=&
3^{-1}2^{2d-2}A_d~m(m^{d-1}C_V)^2|m^{-3/2}\psi(0)|^2,
\end{eqnarray}
with the corresponding branching ratios being
\begin{eqnarray}
\Br(\pP\to\calU_P)&=&\frac{2^{2d-5}}{\pi\alpha^2}A_d
\left(m^{d-1}C_P\right)^2,\\
\Br(\oP\to\calU_V)&=&\frac{3\cdot 2^{2d-6}}{(\pi^2-9)\alpha^3}A_d
\left(m^{d-1}C_V\right)^2.
\end{eqnarray}

The decay rate to a photon plus an unparticle is
\begin{eqnarray*}
d\Gamma=\frac{1}{4m}\left[A_d\theta(p^0)\theta(p^2)(p^2)^{d-2}
\frac{d^4p}{(2\pi)^4}\right]%
\left[\frac{d^3{\bf k}}{(2\pi)^32\omega}\right]
(2\pi)^4\delta^4(p+k-p_1-p_2)|\calA|^2.
\end{eqnarray*}
Upon finishing $p$ integration and for $\calA$ independent of
$|{\bf k}|$ which is the case here, the differential rate in
fractional photon energy and solid angles is
\begin{eqnarray}
\frac{d\Gamma}{dx~d\Omega}=A_d2^{2d-10}\pi^{-3}m^{2d-3}|\calA|^2x(1-x)^{d-2},
\end{eqnarray}
where $p^2\approx 4m(m-\omega)$ is used and the integration region
is fixed by the step functions to be $0\le x\le 1$ with
$x=\omega/m$. Contrary to the particle case where the photon in a
two-body final state is monochromatic, the photon accompanying the
unparticle follows a continuous spectrum. This is again due to the
lack of a mass-shell constraint and the like for unparticles. This
is more than a mere missing energy or momentum that could be used
to separate unparticle signals from ``normal'' new physics.

The unpolarized differential decay rates are, upon finishing the
angular integration,
\begin{eqnarray}
\frac{d\overline{\Gamma}}{dx}
\left(\oP\to\gamma\calU_{S,P,A}\right)&=&\frac{A_d2^{2d-2}}{3\pi}
m|m^{-3/2}\psi(0)|^2\alpha(m^{d-1}C_{S,P,A})^2 ~x(1-x)^{d-2},
\end{eqnarray}
with the total rates and branching ratios being
\begin{eqnarray}
\overline{\Gamma}\left(\oP\to\gamma\calU_{S,P,A}\right)&=&
\frac{4\Gamma\left(d+\frac{1}{2}\right)}{3\Gamma(d+1)\Gamma(2d)}
\pi^{\frac{3}{2}-2d}m|m^{-3/2}\psi(0)|^2\alpha(m^{d-1}C_{S,P,A})^2,\\
\Br\left(\oP\to\gamma\calU_{S,P,A}\right)
&=&\frac{\Gamma\left(d+\frac{1}{2}\right)}{\Gamma(2d)\Gamma(d+1)}
\frac{3}{4(\pi^2-9)\alpha^2}\pi^{\frac{3}{2}-2d}\left(m^{d-1}C_{S,P,A}\right)^2.
\end{eqnarray}

Now we confront our results with data. The most recent measurement
on invisible positronium decays is reported in
Ref.\cite{badertscher}:
\begin{eqnarray}
\Br(\pP\to\textrm{invisible})&\le&4.3\times
10^{-7}~(90\%\textrm{C.L})\label{eqn_invi_p},\\
\Br(\oP\to\textrm{invisible})&\le&4.2\times
10^{-7}~(90\%\textrm{C.L})\label{eqn_invi_o},
\end{eqnarray}
while the most stringent bounds on exotic two-body decays were set
some years ago \cite{asai}:
\begin{eqnarray}
\Br(\oP\to\gamma X^0)\le 1.1\times 10^{-6}~(90\%\textrm{C.L}),
\label{eqn_exotic}
\end{eqnarray}
where $X^0$ is an unknown neutral boson interacting weakly with
ordinary matter. We take $d=\frac{3}{2}$ as previously. Then Eqs.
(\ref{eqn_invi_p},\ref{eqn_invi_o}) imply respectively
\begin{eqnarray}
\Lambda_P\ge 5.6\times 10^2\textrm{ TeV},~~~ \Lambda_V\ge
4.3\times 10^5\textrm{ TeV}. \label{eq_pv}
\end{eqnarray}
Since several channels contribute to the exotic decays, we
consider one unparticle at a time for simplicity, then Eq.
(\ref{eqn_exotic}) gives
\begin{eqnarray}
\Lambda_{S,P,A}\ge 5.1\times 10^2\textrm{ TeV}. \label{eq_spa}
\end{eqnarray}
If we combine the bounds in Eqs. (\ref{eq_pv}) and (\ref{eq_spa}),
the latter is mainly a bound on $\Lambda_{S,A}$. These are more
stringent bounds than those from the electron $g-2$.

Unparticles descending from some high energy scale invariant
theory behave very differently from familiar particles due to the
lack of a mass-shell constraint and a nonintegral scaling
dimension. This makes them phenomenologically very distinctive.
But whether this is observable depends on how feebly they interact
with ordinary matter. We have considered the general effective
interactions of unparticles with the electron, and investigated
their implications on two of the most precisely measured
quantities in QED: the electron $g-2$ and the invisible and exotic
decays of the positronium. We found that the most stringent
constraint is from invisible orthopositronium decays. For a
scaling dimension of $\frac{3}{2}$, the effective energy scale
responsible for the vector unparticle-electron interaction exceeds
$4\times 10^5$ TeV. The bounds on the energy scales of other
unparticles range from a few tens to a few hundreds TeV. This
result makes the experimental observation of unparticles rather
challenging in low energy electron systems. It remains to be seen
whether they are detectable in high energy processes.

%


\end{document}